\documentclass{article}

\usepackage{msarxiv}

\usepackage[utf8]{inputenc} 
\usepackage[T1]{fontenc}    
\usepackage{hyperref}       
\usepackage{url}            
\usepackage{booktabs}       
\usepackage{amsfonts}       
\usepackage{nicefrac}       
\usepackage{microtype}      
\usepackage{lipsum}
\usepackage{graphicx}
\usepackage{booktabs}
\usepackage{amsmath}
\usepackage{longtable}
\usepackage{array}
\usepackage{multirow}
\usepackage{wrapfig}
\usepackage{float}
\usepackage{colortbl}
\usepackage{pdflscape}
\usepackage{tabu}
\usepackage{threeparttable}
\usepackage{threeparttablex}
\usepackage[normalem]{ulem}
\usepackage{xcolor}
\usepackage{authblk}

\title{Combining Cox Regressions Across a Heterogeneous Distributed Research Network Facing Small and Zero Counts}

\author[1,2,3]{Martijn J. Schuemie}
\author[4]{Yong Chen}
\author[1,5]{David Madigan}
\author[1,3,6]{Marc A. Suchard}
\affil[1]{Observational Health Data Sciences and Informatics, New York, New York}
\affil[2]{Janssen Research \& Development, Titusville, NJ}
\affil[3]{Department of Biostatistics, University of Califoria, Los Angeles, CA}
\affil[4]{Department of Biostatistics, Epidemiology and Informatics,University of Pennsylvania, Philadelphia, PA}
\affil[5]{Khoury College of Computer Sciences, Northeastern University, Boston, MA}
\affil[6]{Department of Human Genetics, University of Califoria, Los Angeles, CA}

\begin{document}
\maketitle

\def\tightlist{}

\begin{abstract}
Studies of the effects of medical interventions increasingly take place in distributed research settings using data from multiple clinical data sources including electronic health records and administrative claims. In such settings, privacy concerns typically prohibit sharing of individual patient data, and instead, analyses can only utilize summary statistics from the individual databases. In the specific but very common context of the Cox proportional hazards model, we show that standard meta analysis methods then lead to substantial bias when outcome counts are small. This bias derives primarily from the normal approximations that the methods utilize. Here we propose and evaluate methods that eschew normal approximations in favor of three more flexible approximations: a skew-normal, a one-dimensional grid, and a custom parametric function that mimics the behavior of the Cox likelihood function. In extensive simulation studies we demonstrate how these approximations impact bias in the context of both fixed-effects and (Bayesian) random-effects models. We then apply these approaches to three real-world studies of the comparative safety of antidepressants, each using data from four observational healthcare databases.
\end{abstract}


\hypertarget{intro}{%
\section{Introduction}\label{intro}}

Studies of the effects of medical intervention increasingly take place in distributed research settings using data from multiple clinical data sources.
This is especially true in observational research, where studies draw on existing healthcare data such as electronic health records (EHRs), administrative claims data, and registries.
Such studies provide critical clinical knowledge, especially in settings where randomized trials prove impractical or overly costly.
The recent emergence of distributed research networks, such as the Observational Health Data Sciences and Informatics (OHDSI), (Hripcsak et al. 2016) enables the use of data from hundreds of millions of patients across the world and can answer questions about relationships between exposures and outcomes, even for relatively rare exposures and outcomes.

Despite the promise of multi-site analyses, a number of analytic challenges commonly arise.
First, sharing of individual patient data (IPD) rarely proves possible because of patient privacy concerns and local governance regulations.
Second, the nature of the data gathering process leads to censored observation periods that require a time-to-event analysis rather than simpler incidence rate estimation.
Third, due to the observational nature of the data and the consequent potential for confounding, some correction for baseline differences between exposure groups always proves necessary, typically via stratifying, matching, or weighting by a propensity score (Rosenbaum and Rubin 1983) or disease risk score. (Miettinen 1976)
Fourth, since different data sites represent different patient populations, inter-site heterogeneity often arises.
Finally, even though many of these healthcare databases contain the records of large numbers of patients, co-occurrences of even moderately rare exposures and/or outcomes often prove to be sparse to non-existent.
For example, a recent study comparing the risk of angioedema following exposure to levetiracetam as compared to phenytoin (Duke et al. 2017) used data from 10 databases covering over 300 million patients. The study identified 350,000 patient exposures to one or the other drug but only 125 patients experienced angioedema during exposure and several sites observed zero outcome events during exposure.

To avoid sharing of IPD, studies can at best receive only aggregate statistics from the study sites.
Current practice typically estimates the hazard ratio (with standard error) at each site and then combines these estimates using a traditional random effects meta-analysis. (DerSimonian and Laird 1986)
However, with small counts, the standard normal approximation of the per-site likelihood function can break down and lead to substantial bias,
especially when only one of the treatment arms yields zero outcome events. (Duan et al. 2020)
Recently, with the increasing research interest in rare outcomes in clinical applications such as drug safety, there is a growing body of work addressing the issue of combining small count data from multiple studies.
Much of the literature eschews time-to-event data and concerns itself with sharing and combining 2x2 tables (e.g.~Sweeting, Sutton, and Lambert (2004), Friedrich, Adhikari, and Beyene (2007), Tian et al. (2009), Liu, Liu, and Xie (2014), Cafri et al. (2015), Gronsbell et al. (2020)).
Others have focused specifically on the analysis of time-to-event data but present methods that require sharing of IPD (e.g.~Siannis et al. (2010)).
Some authors have proposed the use of `risk sets,' sets of subjects having similar covariate values who are at risk at various time points. This approach primarily aims to account for heterogeneity in the selection of covariates, and in fact requires the normality assumption to hold even within risk sets. (Yuan and Anderson 2010; Li et al. 2019)

In this paper, we aim to fill a methodological gap on evidence synthesis for time-to-event outcomes with small counts. Specifically, we propose and evaluate methods for combining evidence from Cox models across a distributed research network of databases without sharing IPD.
In Section \ref{likelihoodApproximations}, we propose and investigate four different approximations of the per-site likelihood function: a normal approximation, a skew-normal approximation, a custom parametric approximation, and a grid approximation.
We develop both a fixed-effect and a random-effects meta-analysis model using these approximations (Section \ref{evidenceSynthesis}).
For the latter we build on the work by Seide, Röver, and Friede (2019) and utilize a Bayesian random-effects meta-analysis employing the different approximations of the likelihood.
We evaluate these techniques using extensive simulations (Section \ref{simulationStudies}), and also demonstrate them in three case studies using four real-world databases (Section \ref{appliedExamples}).

\hypertarget{methods}{%
\section{Methods}\label{methods}}

\hypertarget{likelihoodApproximations}{%
\subsection{Likelihood approximations}\label{likelihoodApproximations}}

The key idea of our methods is to directly approximate the shape of the per-site log likelihood function and communicate the parameters that summarize the shape of likelihood function from each site.
For the rest of the paper, we focus on a single parameter Cox proportional hazards model, where the parameter quantifies the treatment effect comparing two exposures. We adjust for confounding variables via a large-scale propensity model using demographics as well as all prior conditions, exposures, procedures, measurements, etc. (Tian et al., 2018).
This single parameter Cox proportional hazard model has proven to be an effective approach in many applications (e.g.~Duke et al. (2017), Suchard et al. (2019), Hripcsak et al. (2020), You et al. (2020)).

We consider four ways to approximate and subsequently communicate the shape of the site-specific likelihood function for our single parameter of interest which we denote \(\beta\).

\hypertarget{normal-approximation}{%
\subsubsection{Normal approximation}\label{normal-approximation}}

Let \(\phi(\beta)\) denote the standard normal probability density function

\[\phi(\beta,\mu,\sigma) = \frac{1}{\sqrt{2\pi\sigma^2}}e^{-\frac{(\beta-\mu)^2}{2\sigma^2}} . \]

We estimate \(\hat{\mu}\) and \(\hat{\sigma}\) using the mode and Fisher information of the Cox partial likelihood function.
Note that if one or both of the treatment groups has zero outcomes, the mode is not defined and the normal approximation does not exist.
Typically, analyses simply remove such sites altogether, potentially biasing the resulting meta-analytic estimate.

\hypertarget{skew-normal-approximation}{%
\subsubsection{Skew-normal approximation}\label{skew-normal-approximation}}

Described by Azzalini (2013), the skew-normal function generalizes the normal distribution and allows for a degree of skewness. The skew-normal density function combines the cumulative normal distribution function:
\[\Phi(\beta,\mu,\sigma) = \int_{-\infty}^{\beta}\phi(t,\mu,\sigma)dt  \]
with the normal density to yield:
\[f(\beta,\mu,\sigma,\alpha) = 2\phi(\beta,\mu,\sigma)\Phi(\alpha \beta,\mu,\sigma) . \]
We estimate \(\hat{\mu}\), \(\hat{\sigma}\), and \(\hat{\alpha}\) using the procedure described below.
Even in the presence of zero counts, the skew-normal can provide a reasonable approximation of the target likelihood function.

\hypertarget{custom-approximation}{%
\subsubsection{Custom approximation}\label{custom-approximation}}

\begin{figure}

{\centering \includegraphics[width=0.8\linewidth]{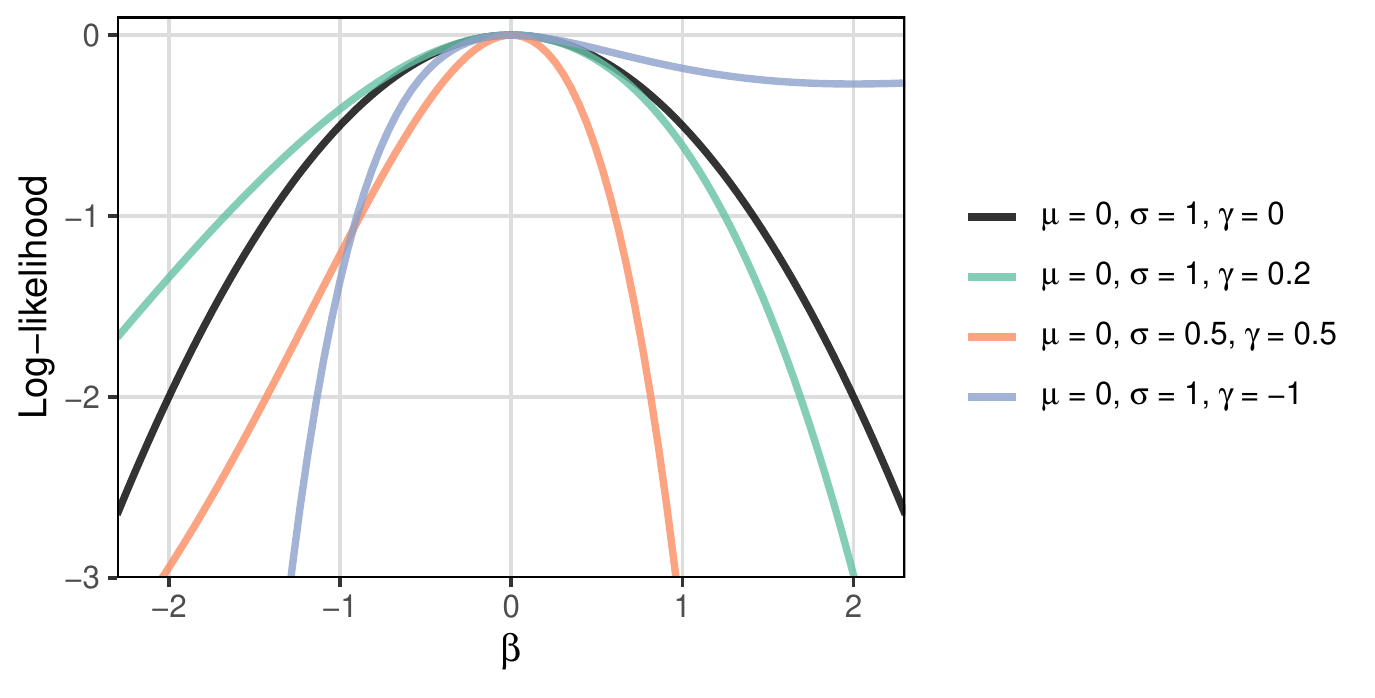}

}

\caption{The custom approximation function, under various parameter choices.}\label{fig:customFunctionPlot}
\end{figure}

Our investigations suggested that the skew-normal performed poorly in the face of severe skewness.
We therefore propose a novel function, which we refer to as the `custom function':
\[ln(f(\beta,\mu,\sigma,\gamma)) = -\frac{(\beta-\mu)^2}{2\sigma^2}e^{\gamma(\beta-\mu)} . \]
We estimate \(\hat{\mu}\), \(\hat{\sigma}\), and \(\hat{\gamma}\) using the procedure described below.
We note that the custom function defaults to a normal when skew is zero (\(\hat{\gamma} = 0\)).
The motivation for this modification of the normal likelihood is to mimic the double exponential component which also features in the Cox likelihood function.
Figure \ref{fig:customFunctionPlot} demonstrates this custom approximation function under various parameter choices.
We have observed this family of functions to perform well even in the presence of zero-counts.

\hypertarget{grid}{%
\subsubsection{Grid}\label{grid}}

Finally, we can communicate the (log) partial likelihood function by sampling values at predefined points in a one-dimensional grid of hazard ratios over a plausible range.
Here we define the grid from a log hazard ratio as 1,000 equally spaced points spanning \(\log(0.1)\) to \(\log(10)\).
Most if not all effect sizes of interest lie well within this range.
We note that zero counts do not impact this approximation and increasing the grid size can provide an arbitrarily high quality approximation.

\hypertarget{function-fitting}{%
\subsection{Function fitting}\label{function-fitting}}

We estimate the parameters for the skew-normal and custom function by minimizing the weighted sum of squares across \(B\), a grid of values for \(\beta\):

\[SS = \sum_{\beta \in B} w(\beta)\{\log[L(\beta)] - \log[f(\beta)]\}^2, \]
where \(L(\beta)\) is the likelihood of \(\beta\), \(f(\beta)\) is our approximation function (skew-normal or custom), and \(w(\beta)\) is the weight.
The weight is simply the likelihood, truncated to a minimum value:
\[w(\beta) = \begin{cases}
        L(\beta) & \text{for } L(\beta) > 10^{-3}, \text{ and}\\
        10^{-3} & \text{ otherwise.}
        \end{cases}\]
We define \(B\) as a predefined grid from \(\log(0.1)\) to \(\log(10)\), in 100 equal steps.

\hypertarget{evidenceSynthesis}{%
\subsection{Evidence synthesis via combining likelihoods from different sites}\label{evidenceSynthesis}}

\hypertarget{fixed-effect-model}{%
\subsubsection{Fixed-effect model}\label{fixed-effect-model}}

A fixed-effect model assumes the true hazard ratio \(\mu\) is the same across all data sites.
One way to estimate \(\hat{\mu}\) is to optimize the overall likelihood, defined by the product of the per-site likelihoods:
\[L_{\text{all}}(\beta) = \prod_{n=1}^NL_n(\beta)\]
where \(N\) is the number of data sites, and \(L_n(\beta)\) is the likelihood of log hazard ratio \(\beta\) at site \(n\).
Here we will use the various approximations to convey \(L_n(\beta)\) (normal, skew-normal, custom, and grid).

Once we have found the optimum, we can find the 95\% confidence interval by finding \(\beta\) with
\[\log[L_{\text{all}}(\hat{\mu})] - \log[L_{\text{all}}(\beta)] = \dfrac{q}{2} , \]
where \(q\) is the 95\% quantile of the \(\chi^2\) distribution with 1 degree of freedom. (Venzon and Moolgavkar 1988)

\hypertarget{random-effect-meta-analysis}{%
\subsubsection{Random-effect meta-analysis}\label{random-effect-meta-analysis}}

A random-effects meta-analysis assumes the true hazard ratio \(\theta_n\) from site \(n\) is drawn from some distribution.
Often, as we will do here, this distribution is assumed to be normal with mean \(\mu\) and variance \(\tau^2\):
\[\theta_n \sim \mathcal{N}(\mu, \tau^2)  . \]
Because we expect to have little statistical power per database, we foresee problems in estimating \(\tau\) accurately. (Friede et al. 2017)
We therefore adopt the Bayesian approach of Seide, Röver, and Friede (2019), estimating \(\hat{\mu}\) as the median of the posterior distribution of \(\mu\).
We identify the 95\% credible interval as the 95\% highest density interval (Kruschke 2011), and compute a proxy of the standard error based on the credible interval and assuming a normal distribution.
Similar to Seide, Röver, and Friede (2019) we use a half-normal prior with scale 0.5 on \(\tau\).
We assume a normal prior with a standard deviation of 2 on \(\mu\), thus specifying 95\% of the probability mass on the hazard ratio is between 0.04 and 26.84.

To learn the joint posterior distribution of \((\mu, \tau^2, \theta_1, \ldots, \theta_N)\), we employ Markov chain Monte Carlo (MCMC) via a random-scan Metropolis-within-Gibbs (Liu, Wong, and Kong 1995) sampling scheme.
The scheme interleaves Gibbs sampling from full conditional distributions \(p(\mu | \tau^2, \theta_1, \ldots, \theta_N)\) and \(p(\tau^2 | \mu, \theta_1, \ldots, \theta_N)\) with separate random-walk Metropolis-Hastings transition kernels on \(\theta_n\) for all \(n\).
These latter kernels have auto-tuning scale constants to efficiently sample from a variety of posteriors.
To approximate each posterior, we simulate MCMC chains for 1.1 million steps, discarding the first 0.1 million steps as burn-in and sub-sample every 100 steps to decrease auto-correlation between samples.
For our applications, these settings generate effective sample sizes of well over 1,000 across all model parameters.
We make our implementation available as an R package \texttt{EvidenceSynthesis} that relies on the popular Bayesian sampling software \texttt{BEAST}. (Suchard et al. 2018)

\hypertarget{simulationStudies}{%
\section{Simulation Studies}\label{simulationStudies}}

\hypertarget{simulation-settings-and-measures-to-compare-performance-of-different-methods}{%
\subsection{Simulation Settings and Measures to compare performance of different methods}\label{simulation-settings-and-measures-to-compare-performance-of-different-methods}}

We perform two separate simulation studies, one assuming fixed effects only, and one where \(\tau\) is allowed to be greater than 0.
The reason for having two separate simulation studies is the computational expense of our random-effects approach, meaning we had to be conservative in the number of simulations.
The simulation framework is described in Appendix A.
We consider the following simulation parameters:

\begin{itemize}
\tightlist
\item
  \textbf{Treated Fraction}: The fraction of the study population treated with the target treatment. The remainder of the population is assumed to have the comparator exposure.
\item
  \textbf{Hazard ratio}: The true (mean) hazard ratio.
\item
  \textbf{N Sites}: The number of database sites.
\item
  \textbf{Max n}: The maximum number of subjects per site (size is sampled from \(unif(1000, Max n)\).
\item
  \textbf{N Strata}: The number of propensity or disease risk score strata. The baseline risk is simulated to be constant within strata but differ across strata.
\item
  \textbf{\(\tau\)}: The true standard deviation of the random effects distribution.
\end{itemize}

For each parameter we select several values, and then create the full factorial combination of all values.
Each unique combination of parameter values forms a `simulation scenario', and is repeated 1,000 times, allowing us to compute the following metrics:

\begin{itemize}
\tightlist
\item
  \textbf{Coverage}: The fraction of times the true hazard ratio is within the 95\% confidence or credible interval.
\item
  \textbf{Bias}: The mean of the difference between the log of the estimated hazard ratio and the log of the true hazard ratio.
\item
  \textbf{MSE}: The mean of the square of the difference between the log of the estimated hazard ratio and the log of the true hazard ratio
\item
  \textbf{Precision}: The geometric mean of the precision (\(1 / (Standard Error)^2\)).
\item
  \textbf{Non-estimable}: The fraction of simulation iterations where an estimate could not be produced, for example because all sites had zero counts
\end{itemize}

\hypertarget{simulation-results}{%
\subsection{Simulation Results}\label{simulation-results}}

In total, we evaluated 540 and 216 unique simulation parameter combinations in the fixed-effects and random-effect simulation studies, respectively, and for each combination we computed the performance metrics.
Our online interactive visualization at \url{https://data.ohdsi.org/NonNormalEvidenceSynthesisSimulations/} enables a complete review of the results.

\hypertarget{fixed-effects-simulations}{%
\subsubsection{Fixed-effects simulations}\label{fixed-effects-simulations}}

\begin{figure}

{\centering \includegraphics[width=1\linewidth]{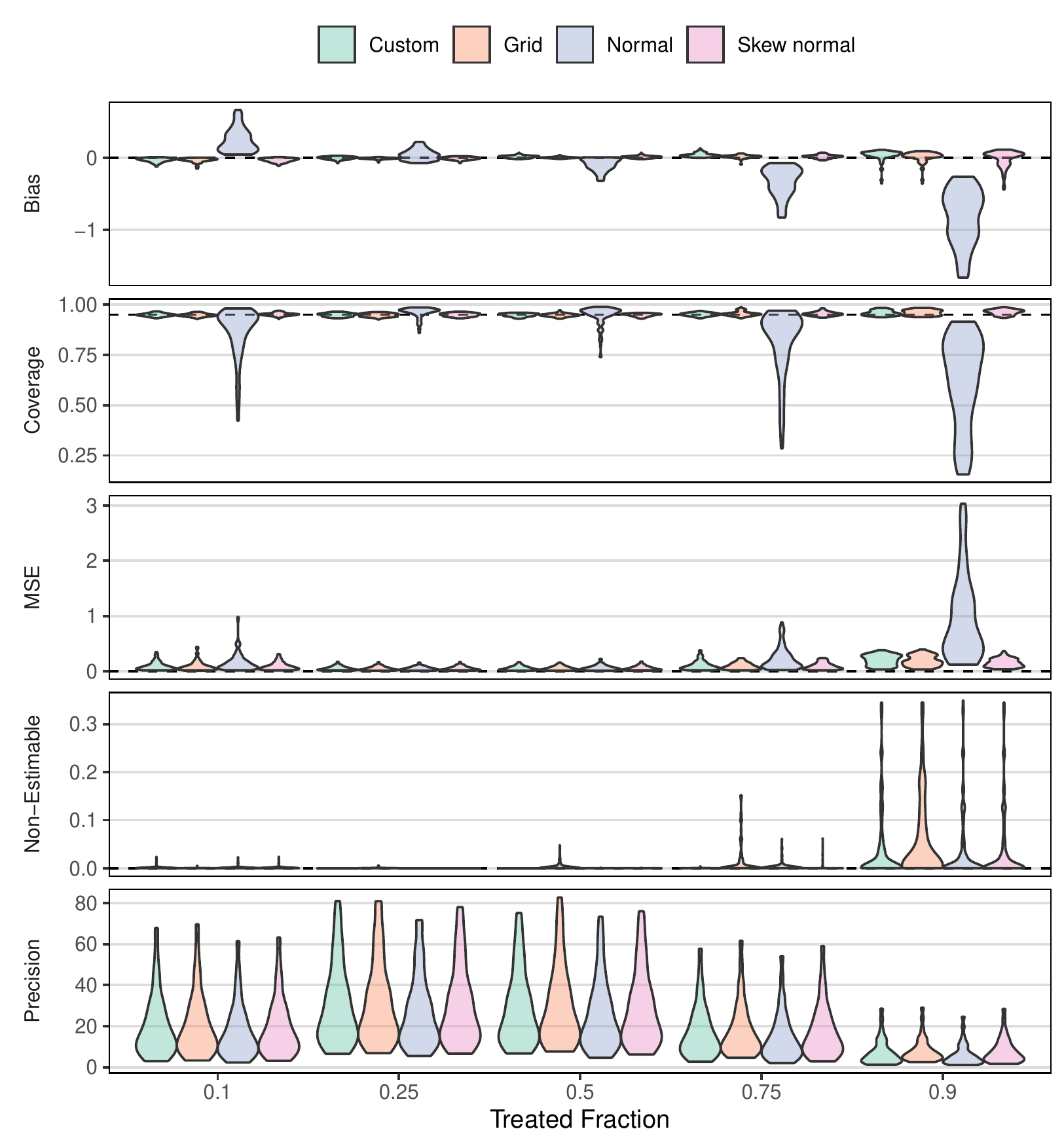}

}

\caption{Distributions of performance metrics of fixed-effects meta-analyses using the various likelihood approximations across the 560 distinct fixed-effects simulation scenarios. The distributions are stratified by the treated fraction, which is one of the simulation parameters that was varied. Dashed lines indicate reference values of a perfect performance, where applicable. MSE = Mean Squared Error.}\label{fig:fixedFxSimulations}
\end{figure}

The violin plots in Figure \ref{fig:fixedFxSimulations} show the distributions of performance metrics across the fixed-effects simulations.
Of the five simulation parameters that we varied, the treatment faction proved to be most important in explaining differences in performance.
That is why we show this parameter on the horizontal axis.
When the treatment groups are of unequal size (i.e.~the treatment fraction moves away from 0.5), the normal approximation shows increasing bias and lower coverage of the 95\% confidence interval.
All other approximations appear essentially unbiased.

\hypertarget{random-effects-simulations}{%
\subsubsection{Random-effects simulations}\label{random-effects-simulations}}

\begin{figure}

{\centering \includegraphics[width=1\linewidth]{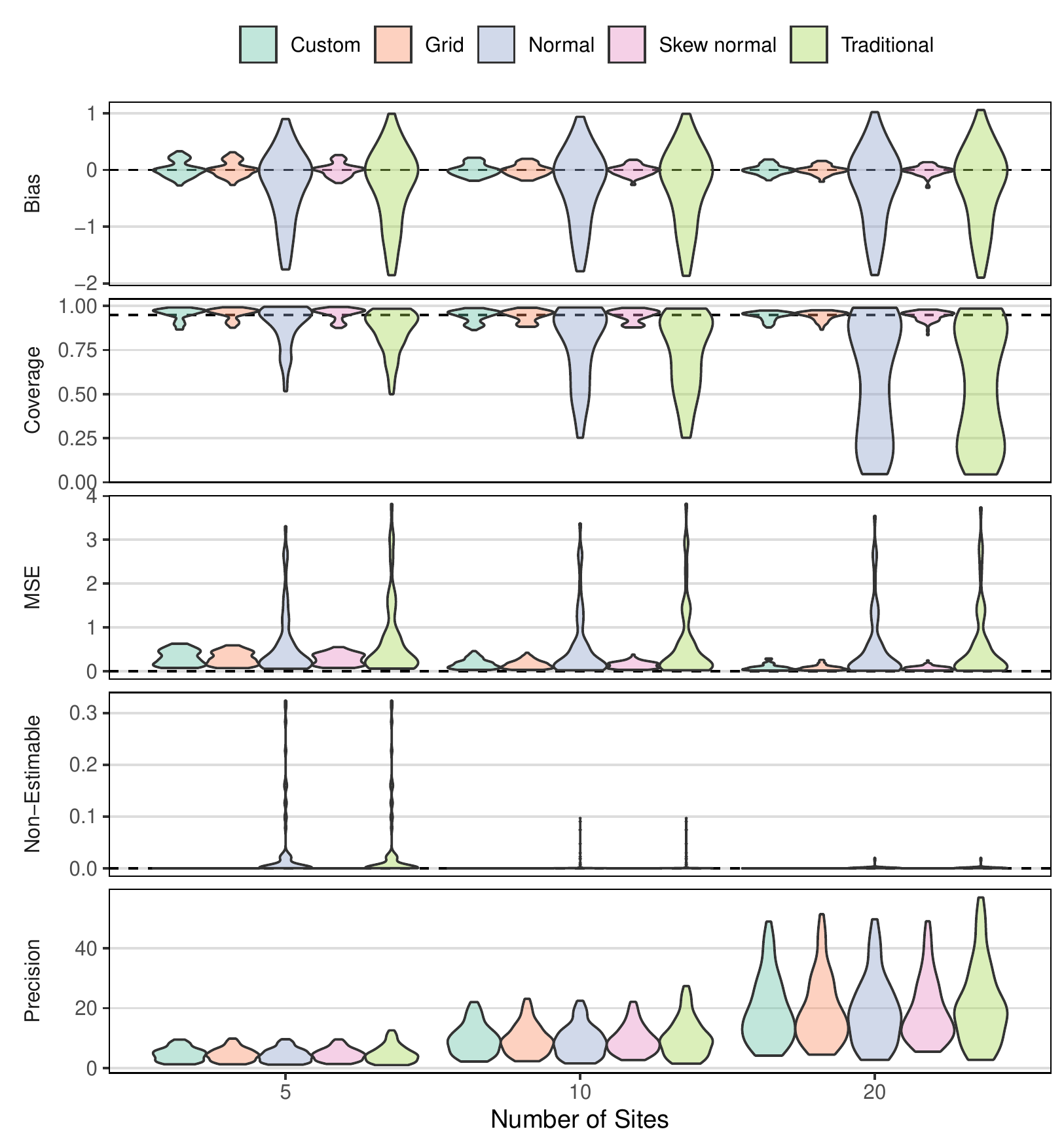}

}

\caption{Distributions of performance metrics of random-effects meta-analyses using the various likelihood approximations across the 216 distinct random-effects simulation scenarios. The distributions are stratified by the number of sites, which is one of the simulation parameters that was varied. Dashed lines indicate reference values of a perfect performance, where applicable. All meta-anlyses were Bayesian, except the "traditional" meta-analyses that used the traditional frequentists random-effects meta-analysis, using a normal approximation. MSE = Mean Squared Error.}\label{fig:randomFxSimulations}
\end{figure}

Although in the random-effects meta-analyses the treated fraction was still an important determinant of performance, we chose to highlight the number of sites in Figure \ref{fig:randomFxSimulations}.
This demonstrates that the low coverage for the normal approximation decreases further as the number of sites increases.
One can also see that the non-normal approximations remain unbiased at all simulation parameter values.

\hypertarget{custom-versus-grid-approximation}{%
\subsubsection{Custom versus grid approximation}\label{custom-versus-grid-approximation}}

\begin{table}

\caption{\label{tab:customGrid}Comparison of performance metrics between the custom and grid approximations in the fixed-effects simulations using a fixed-effects model, and in the random-effects simulations using a random-effects model. The p-value corresponds to a two-sided t-test for paired samples.}
\centering
\begin{tabular}[t]{lrrrrr}
\toprule
Type & Metric & Mean (Custom) & Mean (Grid) & Mean diff. & p\\
\midrule
Fixed-effects & Bias & 0.013 & 0.000 & 0.013 & $<$0.001\\
Fixed-effects & Coverage & 0.951 & 0.952 & -0.000 & 0.085\\
Fixed-effects & MSE & 0.101 & 0.096 & 0.005 & $<$0.001\\
Fixed-effects & Non-Estimable & 0.008 & 0.015 & -0.007 & $<$0.001\\
Fixed-effects & Precision & 20.725 & 21.640 & -0.915 & $<$0.001\\
\addlinespace
Random-effects & Bias & 0.005 & 0.000 & 0.004 & $<$0.001\\
Random-effects & Coverage & 0.946 & 0.948 & -0.002 & $<$0.001\\
Random-effects & MSE & 0.190 & 0.178 & 0.012 & $<$0.001\\
Random-effects & Non-Estimable & 0.000 & 0.000 & 0.000 & \\
Random-effects & Precision & 11.125 & 11.126 & -0.001 & 0.975\\
\bottomrule
\end{tabular}
\end{table}

We conclude that the custom and grid approximations both represent good candidates to minimize bias.
Table \ref{tab:customGrid} explores how these two approaches compare, showing the means and mean differences for the various metrics.
We see the grid approximation achieves slightly lower bias and MSE in all simulations, and slightly higher precision when using a fixed-effects model.
Coverage is largely comparable.

\hypertarget{appliedExamples}{%
\section{Applied examples}\label{appliedExamples}}

\hypertarget{study-design-and-data}{%
\subsection{Study design and data}\label{study-design-and-data}}

Although regulators require all medical treatments to undergo extensive evaluation in clinical trials prior to marketing, ensuring safety post marketing remains a significant societal priority; rare adverse effects may fail to appear in typically-sized clinical trials but result in a significant post-marketing public health challenge.
Large scale, routinely collected healthcare data can shed light on previously unknown side effects, including rare outcomes.
For example, in a prior large-scale study (Schuemie et al. 2018) we used four US administrative claims database to estimate the effect of depression treatments, comparing 17 treatments for 22 outcomes of interest.
From this study we picked three examples to reflect various situations where we encountered small and zero counts:

\begin{itemize}
\tightlist
\item
  \textbf{Example 1:} Amitriptyline (target) vs.~citalopram (comparator) for the risk of acute liver injury. We selected this scenario as it had non-zero counts across all four databases, and the highest mean skew (here interpreted at the ratio between the upper part and the lower part of the 95\% confidence interval).
\item
  \textbf{Example 2:} Nortriptyline (target) vs.~duloxetine (comparator) for the risk of acute liver injury. We selected this scenario as it had two databases with zero counts, and the highest average skew.
\item
  \textbf{Example 3:} Nortriptyline (target) vs.~venlafaxine (comparator) for the risk of decreased libido. We selected this scenario as it had two databases with zero counts in the target, and the highest counts in the comparator.
\end{itemize}

In all three examples we compared new-users of each drug, requiring 365 days of continuous observation prior to treatment initiation as well as a prior diagnose of major depression, but excluding people who had the outcome before treatment initiation.
We define time-at-risk to start on the day of treatment initiation and stop when treatment stopped, allowing for a 30-day gap in treatment continuation.
We identified exposures as any dispensing of a drug containing the mentioned ingredient.
We defined the acute liver injury outcome as a diagnosis of acute liver injury in an emergency room or inpatient setting. (Udo et al. 2016)
We defined decreased libido as any occurrence of a diagnosis indicating decreased libido.
To address confounding, we utilized large-scale propensity models using demographics as well as all prior conditions, exposures, procedures, measurements, etc., and utilized \(L_1\) regularization for the estimation. (Tian, Schuemie, and Suchard 2018)
We then used these propensity scores to stratify the target and comparator cohorts in 10 equally-sized strata, and conditioned the proportional hazards outcome models on those strata.

We used four observational healthcare databases:

\begin{itemize}
\tightlist
\item
  IBM MarketScan\textsuperscript{\textregistered} Commercial Claims and Encounters (CCAE) is an administrative health claims database for active employees, early retirees, COBRA continues, and their dependents insured by employer-sponsored plans (individuals in plans or product lines with fee-for-service plans and fully capitated or partially capitated plans). At the time of analysis CCAE contained 131 million patients.
\item
  IBM MarketScan\textsuperscript{\textregistered} Medicare Supplemental Beneficiaries (MDCR) is an administrative health claims database for Medicare-eligible active and retired employees and their Medicare-eligible dependents from employer-sponsored supplemental plans (predominantly fee-for-service plans). Only plans where both the Medicare-paid amounts and the employer-paid amounts were available and evident on the claims were selected for this database. At the time of analysis MDCR contained 9.6 million patients.
\item
  IBM MarketScan\textsuperscript{\textregistered} Multi-state Medicaid (MDCD) is an administrative health claims database for the pooled healthcare experience of Medicaid enrollees from multiple states. At the time of analysis MDCD contained 21.6 million patients.
\item
  Optum\textsuperscript{\textregistered} De-Identified Clinformatics\textsuperscript{\textregistered} Data Mart Database (Optum) is an adjudicated administrative health claims database for members with private health insurance, who are fully insured in commercial plans or in administrative services only, Legacy Medicare Choice Lives (prior to January 2006), and Medicare Advantage.At the time of analysis Optum contained 74.7 million patients.
\end{itemize}

Schuemie et al. (2018) provides full details of this study including the protocol.

Because we have direct access to all four databases, we were able to pool the IPD extracted from these databases in a single regression model to provide a benchmark of the analysis based on pooled data.
In an ordinary distributed network setting this would not be possible.

\hypertarget{results}{%
\subsection{Results}\label{results}}

\begin{figure}

{\centering \includegraphics[width=1\linewidth]{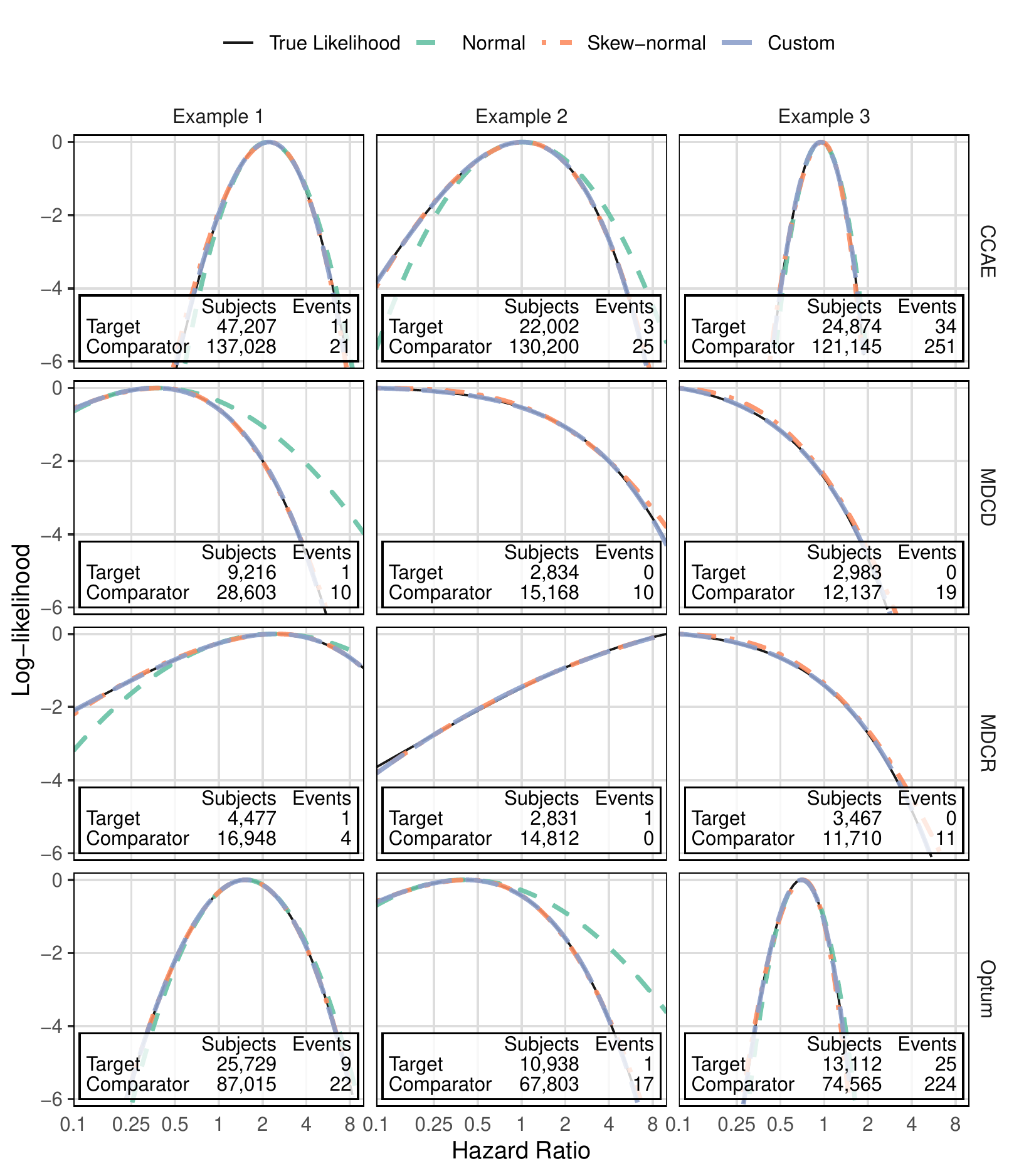}

}

\caption{Three real world examples of evidence synthesis across a network of four databases. For each database, the plot shows the log likelihood function of the Cox regression, as well as the various approximatons. All log likelihood values were normalized so the maximum within the plotted range equals 0. Zero-counts were present for both Example 2 and 3 in the MDCD and MDCR database, where no normal approximation was made. The grid approximation by definition follows the true likelihood in the plotted range, and is therefore not shown.}\label{fig:rwePlot}
\end{figure}

The table insets in Figure \ref{fig:rwePlot} show the number of subjects and outcome events for all three examples in the four databases in the two exposure cohorts.
Although the number of included subjects are often large, the number of events observed during exposure are low, and sometimes zero.
Figure \ref{fig:rwePlot} furthermore shows that in many scenarios the normal approximation provides a poor fit for the actual likelihood functions.
In both Example 2 and 3 in the MDCD and MDCR databases, there is no well-defined maximum likelihood estimate, and no normal approximation exists.
As a consequence, we removed these databases from the meta-analysis for these examples.
In contrast, both the skew-normal and custom approximations hew closely to the true likelihood.
Because we designed the custom approximation to more easily approximate functions with extreme skew, the fitted parameters were smaller than those of the skew-normal.
For example, for Example 2 in the MDCD database, the fitted parameters for the custom approximation were \(\hat{\mu} = -3.72\), \(\hat{\sigma} = 8.39\), and \(\hat{\gamma} = 0.47\), while for skew-normal the fitted parameters were \(\hat{\mu} = 0.27\), \(\hat{\sigma} = 477.64\), and \(\hat{\alpha} = -474.45\).
However, these larger values did not appear to effect the accuracy of the approximation.
Since the grid approximation follows the true likelihood in the plotted range, we do not show it here.

Table \ref{tab:rweTable} shows the hazard ratio estimates generated using the various meta-analysis algorithms.
The gold standard, denoted in bold font, is a Cox regression of the pooled data, which in a real application would not be available because patient-level data cannot be shared.
For Example 1, both the traditional fixed-effects analysis as well as the traditional (non-Bayesian) random-effects meta-analysis (DerSimonian and Laird 1986), both using normal approximations, produce confidence intervals that do not include 1.
However, as the gold standard demonstrates, this statistical significance derives solely and artifactually from the normal approximation.
The normal approximation in a Bayesian random-effects meta-analysis does produce a confidence interval that includes 1, but shifted relative to the gold standard.
In contrast, the skew-normal, custom, and grid approximations all produce meta-analytic estimates close to the gold standard.
For Examples 2 and 3, the normal approximations produce wider confidence intervals than the gold standard and the other approximations, probably because of the two excluded databases.

\begin{table}

\caption{\label{tab:rweTable}Summary hazard ratio estimates (and 95-percent confidence or credible intervals) using the the various evidence synthesis approaches. Bold rows indicate the gold standards using data pooling rather than likelihood approximations in a fixed-effects and Bayesian random-effects model. 'Traditional' analyses are non-Bayesian, and use normal approximations.}
\centering
\begin{tabular}[t]{llll}
\toprule
Method & Example 1 & Example 2 & Example 3\\
\midrule
\textbf{Pooled fixed-effects} & \textbf{1.59 (0.91-2.70)} & \textbf{0.84 (0.28-2.06)} & \textbf{0.76 (0.55-1.03)}\\
Traditional fixed-effects & 1.77 (1.01-3.11) & 0.81 (0.25-2.60) & 0.83 (0.61-1.14)\\
Skew-normal fixed-effects & 1.59 (0.88-2.72) & 0.83 (0.28-2.01) & 0.77 (0.55-1.03)\\
Custom fixed-effects & 1.60 (0.90-2.72) & 0.83 (0.28-2.05) & 0.76 (0.55-1.03)\\
Grid fixed-effects & 1.59 (0.91-2.69) & 0.84 (0.28-2.05) & 0.76 (0.56-1.02)\\
\addlinespace
\textbf{Pooled random-effects} & \textbf{1.45 (0.69-3.00)} & \textbf{0.78 (0.26-2.28)} & \textbf{0.63 (0.28-1.19)}\\
Traditional random-effects & 1.77 (1.01-3.11) & 0.81 (0.25-2.60) & 0.83 (0.61-1.14)\\
Normal random-effects & 1.64 (0.78-3.43) & 0.81 (0.23-2.83) & 0.83 (0.44-1.57)\\
Skew-normal random-effects & 1.43 (0.66-2.96) & 0.77 (0.24-2.26) & 0.64 (0.29-1.20)\\
Custom random-effects & 1.44 (0.67-2.92) & 0.78 (0.24-2.36) & 0.64 (0.28-1.18)\\
Grid random-effects & 1.43 (0.67-2.94) & 0.79 (0.24-2.39) & 0.63 (0.28-1.18)\\
\bottomrule
\end{tabular}
\end{table}

These examples illustrate the profound clinical implications involved with these analytic choices.
For our first example, the traditional approach incorrectly yields a statistically significant effect, potentially causing unwarranted concerns over the safety of amitriptyline.
For the other two examples, more uncertainty would have remained concerning the potential magnitude of the effect size.
Without sharing IPD, the various non-normal approximations were able to produce estimates in line with a gold standard that would not be available in most real-world studies.

\hypertarget{discussion}{%
\section{Discussion}\label{discussion}}

Our simulations show that the use of traditional meta-analytic techniques in distributed research networks can lead to substantial bias and low confidence interval coverage, especially when the compared treatment groups differ in size, and when the number of sites is large (\(n > 10\)).
The clinical applications we presented confirm that this issue exists in real clinical studies, albeit in a less extreme than what the simulation studies suggest.
Since our clinical applications necessarily included a relatively small numbers of sites, we draw little comfort from this attenuation; clinical studies of the future will draw an ever-increasing numbers of sites and standard meta-analytic approximations have great potential to provide misleading results.

We found that combining appropriate approximations of the per-database partial likelihood can effectively eliminate the bias from traditional meta-analysis facing small or zero counts. Such a strategy avoids the sharing of IPD, and hence is suitable for multi-site collaborations, especially for settings such as OHDSI.
The skew-normal, custom, and grid approximation show good performance, in both fixed and random-effects settings.
We believe all three could be used in practice, although the custom approximation may provide the most compact representation to be communicated between sites.
The normal approximation, as suspected, performed poorly in various simulated and real scenarios.

The non-normal approximations evaluated here can account for two situations where the normal approximation fails. First, when counts are low (but not zero), the likelihood will have a maximum but may be skewed, and second, when counts are zero and no maximum exists, but information about the parameter of interest still exists.
Our approximations capture the marginal likelihood for the parameter of interest, typically the effect of the exposure on the risk of the outcome.
The regression model itself can be as complex as it needs to be, in our case, for example, using stratification on a propensity score, or including additional covariates that are not of interest, but adjust for confounding.
We currently specifically focus on Cox regression.
However, this same procedure may also be useful for other types of models, including Poisson and logistic regression.

In this paper we considered hazard ratios in the 0.1 - 10 range as we believe this is realistic.
However, the model can easily extend to different ranges as necessary.

Our current work focuses on the one-dimensional likelihood function.
After proper propensity score stratification, this approach has utility in many pharmacoepidemiological settings.
Future research may explore higher-dimensional approximations, although this may prove intractable with increasing dimensionality.
An alternative direction could be to extend our prior work fitting multi-variable time-to-event models in a distributed settings, which restricts to a neighborhood of a specific point on the likelihood curve. (Duan et al. 2020)

We believe using a non-normal approximation is advisable when performing time-to-event analyses in distributed research settings.
To support this practice and perform the requisite calculations, we have created the \emph{EvidenceSynthesis} R package, which is freely available on the Comprehensive R Archive Network (CRAN).

\hypertarget{acknowledgements}{%
\section{Acknowledgements}\label{acknowledgements}}

This work is partially supported through National Institutes of Health grant U19AI135995 and Food and Drug Administration contract ``OHDSI-based FDA BEST Community Engagement and Development Coordination Center.''

\hypertarget{orcid}{%
\section{ORCID}\label{orcid}}

\noindent
\emph{Martijn Schuemie} \url{http://orcid.org/0000-0002-0817-5361}

\noindent
\emph{Yong Chen} \url{http://orcid.org/0000-0003-0835-0788}

\noindent
\emph{David Madigan} \url{https://orcid.org/0000-0001-9754-1011}

\noindent
\emph{Marc A. Suchard} \url{http://orcid.org/0000-0001-9818-479X}

\hypertarget{supporting-information}{%
\section{Supporting information}\label{supporting-information}}

The reader may explore the full simulation results at \url{https://data.ohdsi.org/NonNormalEvidenceSynthesisSimulations/}.
The description of the simulation studies is available with this paper at the Biometrics web-site on Wiley Online Library.
The R package \emph{EvidenceSynthesis} is available for download on CRAN and contains a vignette illustrating its use.

\hypertarget{references}{%
\section{References}\label{references}}

\indent

\hypertarget{refs}{}
\leavevmode\hypertarget{ref-azzalini_2013}{}%
Azzalini, Adelchi. 2013. \emph{The Skew-Normal and Related Families}. Institute of Mathematical Statistics Monographs. Cambridge University Press. \url{https://doi.org/10.1017/CBO9781139248891}.

\leavevmode\hypertarget{ref-cafri_2015}{}%
Cafri, Guy, Samprit Banerjee, Art Sedrakyan, Liz Paxton, Ove Furnes, Stephen Graves, and Danica Marinac-Dabic. 2015. ``Meta-Analysis of Survival Curve Data Using Distributed Health Data Networks: Application to Hip Arthroplasty Studies of the International Consortium of Orthopaedic Registries.'' \emph{Research Synthesis Methods} 6 (4): 347--56.

\leavevmode\hypertarget{ref-dersimonian_1986}{}%
DerSimonian, Rebecca, and Nan Laird. 1986. ``Meta-Analysis in Clinical Trials.'' \emph{Controlled Clinical Trials} 7 (3): 177--88.

\leavevmode\hypertarget{ref-duan_2020}{}%
Duan, Rui, Chongliang Luo, Martijn J Schuemie, Jiayi Tong, Jason C Liang, Howard H Chang, Mary Regina Boland, et al. 2020. ``Learning from Local to Global-an Efficient Distributed Algorithm for Modeling Time-to-Event Data.'' \emph{bioRxiv}.

\leavevmode\hypertarget{ref-duke_2017}{}%
Duke, Jon D, Patrick B Ryan, Marc A Suchard, George Hripcsak, Peng Jin, Christian Reich, Marie-Sophie Schwalm, et al. 2017. ``Risk of Angioedema Associated with Levetiracetam Compared with Phenytoin: Findings of the Observational Health Data Sciences and Informatics Research Network.'' \emph{Epilepsia} 58 (8): e101--e106.

\leavevmode\hypertarget{ref-friede_2017}{}%
Friede, T., C. Rover, S. Wandel, and B. Neuenschwander. 2017. ``Meta-analysis of few small studies in orphan diseases.'' \emph{Res Synth Methods} 8 (1): 79--91.

\leavevmode\hypertarget{ref-friedrich_2007}{}%
Friedrich, Jan O, Neill KJ Adhikari, and Joseph Beyene. 2007. ``Inclusion of Zero Total Event Trials in Meta-Analyses Maintains Analytic Consistency and Incorporates All Available Data.'' \emph{BMC Medical Research Methodology} 7 (1): 1--6.

\leavevmode\hypertarget{ref-gronsbell_2020}{}%
Gronsbell, Jessica, Chuan Hong, Lei Nie, Ying Lu, and Lu Tian. 2020. ``Exact Inference for the Random-Effect Model for Meta-Analyses with Rare Events.'' \emph{Statistics in Medicine} 39 (3): 252--64.

\leavevmode\hypertarget{ref-hripcsak_2016}{}%
Hripcsak, George, Patrick B Ryan, Jon D Duke, Nigam H Shah, Rae Woong Park, Vojtech Huser, Marc A Suchard, et al. 2016. ``Characterizing Treatment Pathways at Scale Using the Ohdsi Network.'' \emph{Proceedings of the National Academy of Sciences} 113 (27): 7329--36.

\leavevmode\hypertarget{ref-hripcsak_2020}{}%
Hripcsak, George, Marc A Suchard, Steven Shea, RuiJun Chen, Seng Chan You, Nicole Pratt, David Madigan, Harlan M Krumholz, Patrick B Ryan, and Martijn J Schuemie. 2020. ``Comparison of Cardiovascular and Safety Outcomes of Chlorthalidone Vs Hydrochlorothiazide to Treat Hypertension.'' \emph{JAMA Internal Medicine} 180 (4): 542--51.

\leavevmode\hypertarget{ref-kruschke_2011}{}%
Kruschke, John K. 2011. \emph{Doing Bayesian Data Analysis : A Tutorial with R and Bugs}. Burlington, MA: Academic Press. \url{http://www.amazon.com/Doing-Bayesian-Data-Analysis-Tutorial/dp/0123814855}.

\leavevmode\hypertarget{ref-li_2019}{}%
Li, Xiaojuan, Bruce H Fireman, Jeffrey R Curtis, David E Arterburn, David P Fisher, Érick Moyneur, Mia Gallagher, et al. 2019. ``Validity of Privacy-Protecting Analytical Methods That Use Only Aggregate-Level Information to Conduct Multivariable-Adjusted Analysis in Distributed Data Networks.'' \emph{American Journal of Epidemiology} 188 (4): 709--23.

\leavevmode\hypertarget{ref-liu_2014}{}%
Liu, Dungang, Regina Y Liu, and Min-ge Xie. 2014. ``Exact Meta-Analysis Approach for Discrete Data and Its Application to 2x2 Tables with Rare Events.'' \emph{Journal of the American Statistical Association} 109 (508): 1450--65.

\leavevmode\hypertarget{ref-liu1995covariance}{}%
Liu, Jun S, Wing H Wong, and Augustine Kong. 1995. ``Covariance Structure and Convergence Rate of the Gibbs Sampler with Various Scans.'' \emph{Journal of the Royal Statistical Society: Series B (Methodological)} 57 (1): 157--69.

\leavevmode\hypertarget{ref-miettinen_1976}{}%
Miettinen, Olli S. 1976. ``Stratification by a Multivariate Confounder Score.'' \emph{American Journal of Epidemiology} 104 (6): 609--20.

\leavevmode\hypertarget{ref-rosenbaum_1983}{}%
Rosenbaum, P., and D. Rubin. 1983. ``The Central Role of the Propensity Score in Observational Studies for Causal Effects.'' \emph{Biometrika} 70 (April): 41--55. \url{https://doi.org/10.1093/biomet/70.1.41}.

\leavevmode\hypertarget{ref-schuemie_2018}{}%
Schuemie, M. J., P. B. Ryan, G. Hripcsak, D. Madigan, and M. A. Suchard. 2018. ``Improving reproducibility by using high-throughput observational studies with empirical calibration.'' \emph{Philos Trans A Math Phys Eng Sci} 376 (2128).

\leavevmode\hypertarget{ref-seide_2019}{}%
Seide, S. E., C. Röver, and T. Friede. 2019. ``Likelihood-based random-effects meta-analysis with few studies: empirical and simulation studies.'' \emph{BMC Med Res Methodol} 19 (1): 16.

\leavevmode\hypertarget{ref-siannis_2010}{}%
Siannis, F, JK Barrett, VT Farewell, and JF Tierney. 2010. ``One-Stage Parametric Meta-Analysis of Time-to-Event Outcomes.'' \emph{Statistics in Medicine} 29 (29): 3030--45.

\leavevmode\hypertarget{ref-suchard2018bayesian}{}%
Suchard, Marc A, Philippe Lemey, Guy Baele, Daniel L Ayres, Alexei J Drummond, and Andrew Rambaut. 2018. ``Bayesian Phylogenetic and Phylodynamic Data Integration Using Beast 1.10.'' \emph{Virus Evolution} 4 (1): vey016.

\leavevmode\hypertarget{ref-suchard_2019}{}%
Suchard, Marc A, Martijn J Schuemie, Harlan M Krumholz, Seng Chan You, RuiJun Chen, Nicole Pratt, Christian G Reich, et al. 2019. ``Comprehensive Comparative Effectiveness and Safety of First-Line Antihypertensive Drug Classes: A Systematic, Multinational, Large-Scale Analysis.'' \emph{The Lancet} 394 (10211): 1816--26.

\leavevmode\hypertarget{ref-sweeting_2004}{}%
Sweeting, Michael J., Alexander J. Sutton, and Paul C. Lambert. 2004. ``What to Add to Nothing? Use and Avoidance of Continuity Corrections in Meta-Analysis of Sparse Data.'' \emph{Statistics in Medicine} 23 (9): 1351--75.

\leavevmode\hypertarget{ref-tian_2009}{}%
Tian, Lu, Tianxi Cai, Marc A Pfeffer, Nikita Piankov, Pierre-Yves Cremieux, and LJ Wei. 2009. ``Exact and Efficient Inference Procedure for Meta-Analysis and Its Application to the Analysis of Independent 2\(\times\) 2 Tables with All Available Data but Without Artificial Continuity Correction.'' \emph{Biostatistics} 10 (2): 275--81.

\leavevmode\hypertarget{ref-tian_2018}{}%
Tian, Yuxi, Martijn J Schuemie, and Marc A Suchard. 2018. ``Evaluating Large-Scale Propensity Score Performance Through Real-World and Synthetic Data Experiments.'' \emph{International Journal of Epidemiology} 47 (6): 2005--14.

\leavevmode\hypertarget{ref-udo_2016}{}%
Udo, R., A. H. Maitland-van der Zee, T. C. Egberts, J. H. den Breeijen, H. G. Leufkens, W. W. van Solinge, and M. L. De Bruin. 2016. ``Validity of diagnostic codes and laboratory measurements to identify patients with idiopathic acute liver injury in a hospital database.'' \emph{Pharmacoepidemiol Drug Saf} 25 Suppl 1 (March): 21--28.

\leavevmode\hypertarget{ref-venzon_1988}{}%
Venzon, DJ, and SH Moolgavkar. 1988. ``A Method for Computing Profile-Likelihood-Based Confidence Intervals.'' \emph{Journal of the Royal Statistical Society: Series C (Applied Statistics)} 37 (1): 87--94.

\leavevmode\hypertarget{ref-you_2020}{}%
You, S. C., Y. Rho, B. Bikdeli, J. Kim, A. Siapos, J. Weaver, A. Londhe, et al. 2020. ``Association of Ticagrelor Vs Clopidogrel With Net Adverse Clinical Events in Patients With Acute Coronary Syndrome Undergoing Percutaneous Coronary Intervention.'' \emph{JAMA} 324 (16): 1640--50.

\leavevmode\hypertarget{ref-yuan_2010}{}%
Yuan, Xing, and Stewart J Anderson. 2010. ``Meta-Analysis Methodology for Combining Treatment Effects from Cox Proportional Hazard Models with Different Covariate Adjustments.'' \emph{Biometrical Journal} 52 (4): 519--37.

\bibliographystyle{unsrt}
\bibliography{Manuscript.bib}

\end{document}